# The Maunakea Spectroscopic Explorer

Submitted to the Astro2020 call for APC white papers

**White Paper Description:**
The Maunakea Spectroscopic Explorer is a next-generation massively multiplexed spectroscopic facility currently under development in Hawaii. It is completely dedicated to large-scale spectroscopic surveys and will enable transformative science. In this white paper we summarize the science case and describe the current state of the project.

**Corresponding author:**
Jennifer Marshall, MSE/Texas A&M University, marshall@mse.cfht.hawaii.edu

**Co-authors:**
Adam Bolton, NOAO
James Bullock, UC Irvine
Adam Burgasser, UC San Diego
Ken Chambers, University of Hawai'i
Darren DePoy, Texas A&M University
Arjun Dey, NOAO
Nicolas Flagey, MSE
Alexis Hill, MSE
Lynne Hillenbrand, Caltech
Daniel Huber, University of Hawai'i
Ting Li, Fermilab
Stephanie Juneau, NOAO
Manoj Kaplinghat, University of California Irvine
Mario Mateo, University of Michigan
Alan McConnachie, MSE
Jeffrey Newman, University of Pittsburgh
Andreea Petric, MSE/University of Hawai'i
David Schlegel, Lawrence Berkeley National Laboratory
Andrew Sheinis, MSE/CFHT
Yue Shen, University of Illinois at Urbana-Champaign
Doug Simons, MSE/CFHT
Michael Strauss, Princeton
Kei Szeto, MSE
Kim-Vy Tran, Texas A&M University
Christophe Yèche, IRFU, CEA, Université Paris-Saclay
and the MSE Science Team

**Executive summary**

The Maunakea Spectroscopic Explorer (MSE) is the first of a future generation of massively multiplexed spectroscopic facilities. Building upon the success of imminent survey projects such as SDSS-V, DESI, and PFS, MSE is designed to enable truly transformative science, being completely dedicated to large-scale multi-object spectroscopic surveys, each studying thousands to millions of astrophysical objects. MSE will use an 11.25 m aperture telescope to feed 4,332 fibers over a 1.5 square degree field of view and has the capability to observe at a range of spectral resolutions, from R~3,000 to R~40,000, with all spectral resolutions available at all times across the entire field. With these capabilities, MSE will collect more than 10 million fiber-hours of 10m-class spectroscopic observations every year and is designed to excel at precision studies of large samples of faint astrophysical targets.

# 1 Introduction

The past two decades have seen major advancements in our understanding of the Universe thanks in large part to modern optical and near-IR wide field imaging surveys such as SDSS (York et al. 2000), 2MASS (Skrutskie et al. 2006), Pan-STARRS (Chambers et al. 2016), the Hyper Suprime-Cam Subaru Strategic Program (Aihara et al. 2018), the Dark Energy Survey (Dark Energy Survey Collaboration 2018), and Gaia (Gaia Collaboration 2016). Soon the Large Synoptic Survey Telescope (LSST; Ivezić et al. 2019) will open another new window on the Universe, enabling the discovery and study of stars and galaxies too faint to be studied previously over enormous solid angles. To realize the full scientific potential of these surveys, many (if not most) of the objects discovered in these imaging surveys must be studied in further detail using spectroscopic techniques; this has been emphatically demonstrated by the continued success of SDSS's combination of an imaging survey with spectroscopic followup, resulting in a project that is still scientifically productive decades after its first light.

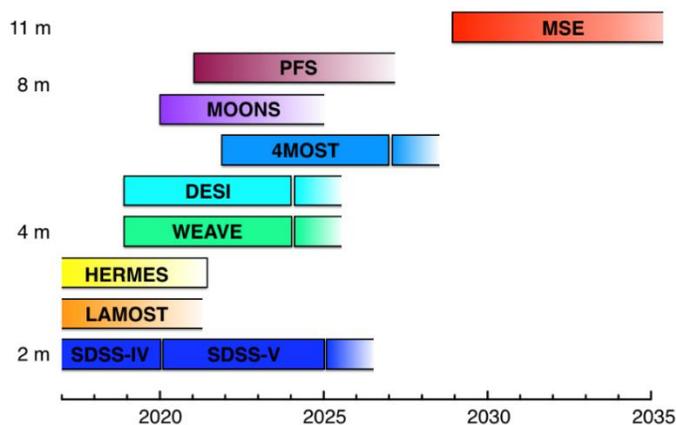

Figure 1: Current anticipated timelines (horizontal axis) for existing and planned massively multiplexed spectroscopic surveys according to telescope aperture (vertical axis). Bounded boxes indicate the duration or lifetime of the survey or facility; absence of a vertical solid line indicates the facility has no clear end date.

The 2010 Decadal Survey of Astronomy produced by the U.S. National Academies identified multi-object spectroscopic instrumentation on large telescopes as an essential capability to be developed over the past decade. Furthermore, in 2015 the National Research Council explicitly emphasized the need to develop new massively multiplexed spectroscopic facilities in order to enhance the scientific return of the LSST project in its report "Optimizing the U.S. Ground-Based Optical and Infrared Astronomy System".  Today the first generation of massively multiplexed spectroscopic facilities are in the planning phase or nearing execution; see Figure 1. These existing and planned facilities go far towards realizing the goals of the past Decadal Survey: SDSS, installed on a 2.5m diameter telescope, can reach $r = 19$ mag, while DESI will go 1-2 mag deeper on a 4m telescope.  PFS will be able to reach 23.5 mag with the 8m Subaru telescope.  It should be noted, however, that none of these facilities combines a collecting area large enough to target the faintest objects discovered by LSST, and none of the US-based facilities is capable of reaching the entire LSST footprint.  As an 11.25m aperture facility, MSE can obtain spectroscopic data for sources identified in a single pass of LSST, down to $r = 24$ mag, and MSE's location allows it to observe more than half of LSST's primary survey area.

## 2   Key science goals and objectives

Over the next decade, these massively multiplexed spectroscopic instruments will enable a plethora of unique, high impact, and exceptionally diverse transformational science cases. Given the large numbers of objects to be studied (some *individual* science cases require observations of tens of millions of objects!), an efficient process must be developed to carry out this science. Observing time, and scientific productivity, will of course be maximized if the spectroscopic surveys are dedicated, or if sufficient time is dedicated to the instruments, and if the capabilities in terms of sensitivity, wavelength coverage, and target density are well-matched to following up the imaging surveys.  In the specific case of MSE, a dedicated, massively multiplexed instrument to be deployed on a 10m-class telescope, the MSE science team has recently produced an up-to-date Detailed Science Case (MSE Science Team 2019) that describes a very large number of science cases, most of which are impossible with current generation instrumentation.  Many compelling, next-generation investigations requiring massively multiplexed spectroscopy, both those in the MSE Detailed Science Case and beyond it, were submitted by members of the MSE Science Team to the Science White Paper call of the Astro2020 Decadal Survey.  Among the huge variety of science questions that MSE can address, we highlight a few of these here:

- **Advancing the knowledge of the nature of dark matter (Li et al. 2019):** Using large-scale precision measurements of the velocities of objects whose motions are affected by gravitational influence of dark matter such as stars, stellar streams, dwarf galaxies in the local Universe, as well as the gravitational lensing systems and galaxy clusters in the distant Universe, MSE will be able to precisely determine the properties of the dark matter halos, including the dark matter subhalo mass function, the phase-space distribution of subhalos, and the dark matter halo density profiles. These observations will constrain the properties of CDM, including testing whether it is purely collisionless.
- **Measuring the mass of the neutrino (Percival et al. 2019):** MSE will produce a high redshift (1.6<z<4.0), large-volume galaxy survey of tens of millions of galaxies.  By measuring the scale-dependent growth of matter fluctuations as traced by galaxies, MSE will produce the first independent 3-σ measurement of the summed neutrino mass.  When combined with similar measurements at different redshifts from DESI and CMBS4, MSE will enable 5-σ measurements of both the summed neutrino mass and the hierarchy of masses.
- **Determining the origin of the elements in the periodic table (Roederer et al. 2019, Johnson et al. 2019):** One-quarter of MSE's 4,332 fibers will feed high resolution (up to R~40,000) spectrographs; with this capability an unprecedentedly large number of detailed stellar chemical abundances will be measured.  Within the Milky Way and Local Group, amassing a large census of normal and peculiar stellar atmospheres, when combined with positional and kinematic information, will enable a significantly more complete study of Galactic archaeology than has been possible to date. In particular, MSE will enable for the first time a full understanding of the formation sites of the *r*-process elements, the only elements on the periodic table whose formation has yet to be explained observationally.

- **Multi-messenger and time domain astronomy:** Time domain astronomy has of late become a focus of observational astronomy: the LIGO-Virgo discovery and localization of neutron star merger GW170817 spectacularly ushered in the era of multi-messenger astrophysics with gravitational waves and has highlighted the immediate need to expand the rapid response capability of follow-up facilities. In the near future, LSST will begin producing overwhelming numbers of transient objects that will require rapid imaging and spectroscopic followup. MSE, with its real-time control software, will enable a new era of rapid-response and time domain astronomy, studying future LIGO events as well as periodic (e.g., binaries/exoplanets, pulsation), evolutionary (e.g., reverberation mapping, Shen et al. 2019) and bursting behavior (e.g., flares, CV novae), on time scales that range from minutes to years.

Additional Science White Papers specifically highlight the need for a facility like MSE. These projects will use the new facility to: spectroscopically investigate young stars (Hillenbrand 2019), ultracool dwarf stars (Burgasser et al. 2019) and evolved stars (Ridgway et al. 2019); investigate the Galactic bulge (Rich et al. 2019) and Andromeda (Gilbert et al. 2019); study Galactic archaeology (Huber et al. 2019) and the assembly of the Milky Way (Sanderson et al. 2019, Dey et al. 2019;, study the ISM (Flagey et al. 2019) and high redshift obscured quasars (Petric et al. 2019); train photometric redshifts of distant galaxies (Newman, 2019); use the synergy between ground and space-based facility to study the Cosmic Dawn (Cuby et al. 2019); and investigate the high redshift Universe to understand inflation (Ferraro et al. 2019). Furthermore, the importance of the "science platform," i.e. the software infrastructure enabling users to efficiently access, analyze, visualize, and model the data, underlying projects like MSE cannot be understated (e.g. Olsen et al. 2019). As a consequence, MSE is envisioned from the start as a full science platform for the design, execution, and exploitation of wide field spectroscopic surveys enabling a vast range of science programs.

As can be seen, the next generation science that not only is enabled by, but indeed, *requires* a dedicated, massively multiplexed spectroscopic facility like MSE spans the entire field of observational astronomy. It is important to stress that these driving science programs *cannot* be addressed by other facilities. For example, no other existing or planned survey spectrographs target the blue/UV at sufficiently high resolution to measure *r*-process and many other elemental abundances, no other facilities include the H-band spectroscopy that is required to explore the full range of physical properties of galaxies that dominate the stellar mass budget at "cosmic noon" ($z\sim2$), and no other facilities have the sensitivity and fiber density to efficiently probe the structure of the Universe with sufficient spatial detail to measure non-gaussianity and neutrino masses. In other cases, it would be impossible to ever acquire sufficient integration times to execute the science program with another non-dedicated instrument, or an instrument installed on a smaller telescope. The currently planned detailed science case for MSE is an impressive compilation of next-generation science themes, but even more impressive is the flexibility that the MSE instrumentation allows for development of many more future science cases.

## 3 Technical overview

The scientific impact of MSE will be made possible and attainable by transforming the existing Canada-France-Hawaii Telescope (CFHT) infrastructure on the Maunakea summit, Hawaii.

CFHT is located at a world-class astronomical site with excellent free-atmosphere seeing (0.4 arcseconds median seeing at 500 nm). The Mauna Kea Science Reserve Comprehensive Management Plan (Kuʻiwalu 2009) for the Astronomy Precinct explicitly recognizes CFHT as one of the sites identified for redevelopment. In order to minimize environmental and cultural impacts to the site, and also to minimize cost, MSE will replace CFHT with an 11.25 m aperture telescope, while retaining the current summit facility. MSE will greatly benefit by building on the technical, community, and cultural experience of CFHT throughout the development of the project.

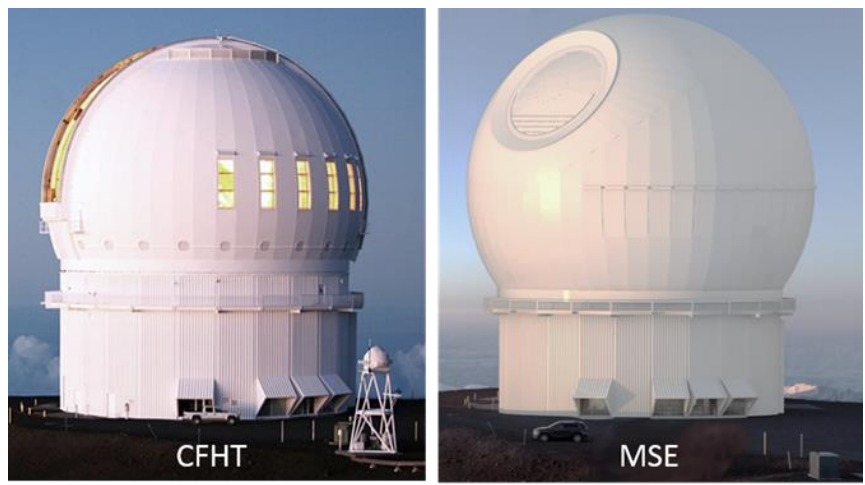

Figure 2: Exterior views of CFHT and MSE.

The rotating CFHT enclosure will be replaced by a Calotte enclosure that is only 10% larger in diameter than the current size, leaving the foundation and much of the current infrastructure intact. The observatory building renovations and structural upgrades will be internal, so the outward appearance of MSE will remain very much unchanged from that of CFHT (Figure 2). Inside, however, a modern observatory will perform cutting-edge science at one of the best astronomical sites in the world, with access to three quarters of the entire night sky.

MSE is designed to take advantage of the excellent site characteristics of Maunakea, which allows for an extremely sensitive, wide-field, and massively multiplexed facility (see Figure 3 and Table 1). At MSE's prime focus, 4,332 input fibers are packed into a hexagonal array. The fibers are positioned to an accuracy of 60 milliarcsec in order to maximize the amount of light injected from science targets into the input fibers, which collect and transmit light to banks of spectrographs tens of meters away. The exquisite seeing allows the fiber diameters to be kept small (85 micron diameter, 0.8 arcseconds, for the high resolution spectrographs and 107 micron diameter, 1.0 arcseconds, for the low and moderate resolution spectrographs), thus keeping the size and cost of the spectrographs attainable. One bank of spectrographs receives light from 3,249 fibers on the focal surface and may be used in either low resolution (R~3,000) or moderate resolution (R~6,000) mode, covering the optical to near-IR wavelength range of 0.36–1.8 microns. Concurrently, the other bank of spectrographs receives light from 1,083 fibers on the focal surface and is dedicated to collecting high resolution spectra in three targeted optical wavelength windows within the wavelength range of 0.36–0.5 microns at R~40,000 and 0.5–0.9 microns at R~20,000. All resolution modes have simultaneous full field coverage of 1.5 swuare degrees, and the massive multiplexing results in the ability to collect many thousands of spectra per hour and over a million spectra per month, all of which will be made available to the MSE user community. Moreover, an upgrade path to add an Integral Field Unit (IFU) system has been incorporated into the design as a second-generation capability for MSE.

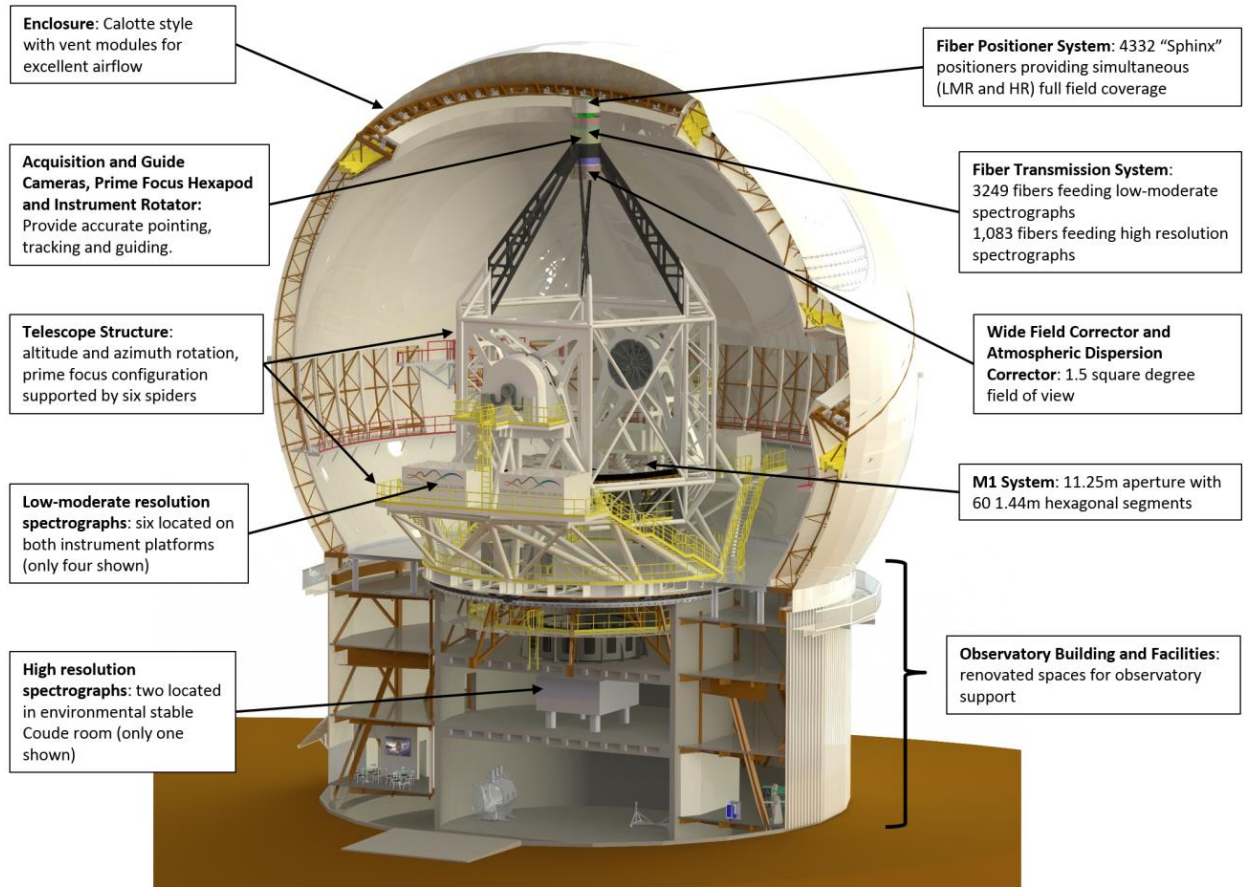

Figure 3: MSE Observatory architecture.

Aside from the physical infrastructure, MSE's success is enabled by efficiently scheduled and executed surveys, by the quality of the data collected, and by MSE's ability to make the science products available to survey teams in a timely and efficient manner. MSE will devote approximately 80% of available time to executing large, homogeneous legacy surveys which will typically require several years to complete. More focused strategic programs, which require smaller amounts of observing time and typically lead to more rapid publications, will occupy the remaining 20% of observing time. Proposals for both types of programs will be solicited from the MSE user community at regular intervals. MSE is operated solely in a queue-based mode, enabled by sophisticated scheduling software. Data are made available to the survey team immediately, and to the whole MSE community on a short timescale. All data and derived products will have a relatively brief proprietary period but will eventually become public to the entire astronomical community.

Table 1: The detailed science capabilities of MSE.

| Site characteristics | |
|---|---|
| Observatory latitude | 19.9 degrees |
| Accessible Sky | 30,000 square degrees (airmass < 1.55 i.e., δ > -30 degrees) |
| Median image quality | 0.37 arcsec (free atmosphere, zenith, 500 nm) |
| Average length of night | 10.2 hours |
| Historical weather losses (average) | 2.2 hours / night |
| Observing efficiency (on-sky, on-target) | 80% |
| Expected on-target science observing hours | 2336 hours / year |
| Expected on-target fiber-hours | 10,112,544 fiber-hours / year (total): 7,589,664 (LR & MR) / 2,529,888 (HR) |

| Telescope architecture | |
|---|---|
| Structure, focus | Altitude-azimuth, Prime |
| M1 aperture | 80.8 m$^2$ |
| Science field of view | 1.52 square degrees |
| Spectrograph system | 6 x LMR spectrographs (4 channels/spectrogrpah, all identical, each channel seperately switchable to provide LR and MR modes) 2 x HR spectrographs (3 channels/spectrograph), both identical, to provide high resolution mode. All spectrographs always available with full multiplexing. Deployable IFU system using LR /MR spectrograph system available as second generation capability |

| Fiber positioning system | |
|---|---|
| Multiplexing | 4,329 (total): 3,249 (LR & MR) / 1,083 (HR) |
| Fiber size | 1 arcsec (LR & MR) / 0.8 arcsec (HR) |
| Positioner patrol radius | 90.3 arcsecs |
| Positioner accuracy | 0.06 arcsec rms |
| Positioner closest approach | Two fibers can approach with 7 arcsecs of each other (three fibers can be placed within 9.9 arcsec diameter circle) |
| Repositioning time | < 120 seconds |
| Typical allocation efficiency | > 80 % (assuming source density approximately matched to fiber density) |

| Low resolution (LR) spectroscopy | | | | |
|---|---|---|---|---|
| Wavelength range | 360 ≦ λ ≦ 560 nm | 540 ≦ λ ≦ 740 nm | 715 ≦ λ ≦ 985 nm | 960 ≦ λ ≦ 1320 nm |
| Spectral resolution (approx. at center of band) | 2,550 | 3,650 | 3,600 | 3,600 |
| Sensitivity requirement (pt. source, 1hr, zenith, median seeing, monochromatic magnitude) | m = 24.0 SNR/res. elem. = 2, λ > 400 nm SNR/res. elem. = 1, λ ≦ 400 nm | m = 24.0 SNR/resolution element = 2 | m = 24.0 SNR/resolution element = 2 | m = 24.0 SNR/resolution element = 2 |

| Moderate resolution (MR) spectroscopy | | | | |
|---|---|---|---|---|
| Wavelength range | 391 ≦ λ ≦ 510 nm | 576 ≦ λ ≦ 700 nm | 737 ≦ λ ≦ 900 nm | 1457 ≦ λ ≦ 1780 nm |
| Spectral resolution (approx. at center of band) | 4,400 | 6,200 | 6,100 | 6,000 |
| Sensitivity requirement (pt. source, 1hr, zenith, median seeing, monochromatic magnitude) | m = 23.5 SNR/res. elem. = 2, λ > 400 nm SNR/res. elem. = 1, λ ≦ 400 nm | m = 23.5 SNR/resolution element = 2 | m = 23.5 SNR/resolution element = 2 | m = 24.0 SNR/resolution element = 2 |

| High resolution (HR) spectroscopy | | | |
|---|---|---|---|
| Wavelength range | 360 ≦ λ ≦ 460 nm | 440 ≦ λ ≦ 620 nm | 600 ≦ λ ≦ 900 nm |
| Wavelength band | λ / 30 [ baseline: 401 - 415 nm ] | λ / 30 [ baseline: 472 - 488.5 nm ] | λ / 15 [ baseline: 626.5 - 672 nm ] |
| Spectral resolution (approx. at center of band) | 40,000 | 40,000 | 20,000 |
| Sensitivity requirement (pt. source, 1hr, zenith, median seeing, monochromatic magnitude) | m = 20.0 SNR/resolution element = 10, λ > 400 nm SNR/resolution element = 5, λ ≦ 400 nm | m = 20.0 SNR/resolution element = 10 | m = 24.0 SNR/resolution element = 10 |

| Science calibration | |
|---|---|
| Sky subtraction accuracy | 0.5% requirement (0.1% goal) |
| Velocity precision | 100 m/s (HR, SNR/resolution element = 30) |
| Relative spectrophotometric accuracy | 3% (LR, SNR/resolution element = 30) |

## 4  Organization, partnership, and current status

MSE was conceived in the early 2010s as a reimagining of the CFHT observatory and has steadily progressed through significant development over the past decade. The MSE Project Office is located at CFHT headquarters in Waimea, HI. In addition to CFHT members Canada, France, and the University of Hawaii, MSE partners include institutes in the countries of Australia, China, and India, as well as "observer" partners NOAO, Texas A&M University, and a large consortium of UK universities, who intend to join as full partners in the near future. Many additional US and international institutions have recently expressed interest in joining the project in the future as well. A significant US-based partnership in the project makes sense not only because the project is located in the US, but also because fully one-quarter of the nearly 400-person international science team are US scientists.

MSE began its Conceptual Design Phase in 2016 (Murowinski et al. 2016; Szeto et al. 2016),

and after a series of subsystem reviews in 2017, the MSE Conceptual Design Phase culminated in January 2018 with a system design review to determine whether the baseline design would meet the science requirements. The external review committee reviewed the project very favorably, approving a transition from Conceptual Design to Preliminary Design phase, and provided valuable suggestions for the project to follow as it advances.

Most recently, the Project Office has worked to coordinate the efforts of the Science Team, with members in 30 countries, to produce an updated Detailed Science Case to refresh the 5-year old original MSE science case. Through this process the number of Science Team members nearly quadrupled, adding scientists from a wide range of US and international institutions and comprising diverse scientific interests. The next phases of the project are to use the new Detailed Science Case to produce a Design Reference Survey, providing a detailed plan for the observations of the first five years of MSE's survey programs, one of the primary recommendations of the Conceptual Design review panel. In parallel, the 100-member international MSE engineering team, composed of instrument scientists, engineers, and technicians from institutes within the current MSE partners has already begun to advance selected subsystems through the Preliminary Design Phase. The MSE project plans to carry out this next phase by partnering with additional US and international instrumentation groups who are committed to take on these design and build efforts in exchange for access to the MSE data products. This model has worked well in the past for projects like SDSS, DES, DESI, 4MOST and others; MSE is poised to follow their lead.

## 5 Project schedule

After completing the Conceptual Design Phase in 2018, the remaining project development schedule is 10 years. The operational lifetime of MSE is designed for 35 years, at the start of science operations. Figure 4 shows the timeline of the project development schedule with the science operations starting in 2030. This timescale is well-suited to the renegotiation of the new Maunakea Master Lease process that is currently underway so that agreement may be made well in advance of the Lease expiring in 2033.

The current timeline is organized into four major activity blocks beginning with the Preliminary Design Phase in 2019, and followed by the overlapping activities (starting in mid-2021) of Detailed Design Phase; Manufacturing and Testing of Subsystems; Assembly, Integration and Verification (AIV) of Subsystems; and Science Commission. The Industrial Subsystems are the observatory building, enclosure and telescope mount, and the Science Instrument Package contains subsystems of robotic positioners, optical fiber bundles, spectrographs and science calibration hardware. Not shown in Figure 4 but intrinsic to the AIV activities are the validation of the observatory control system and the observing program execution software system that schedules, collects, reduces, archives and distributes the final science data products as the MSE science platform.

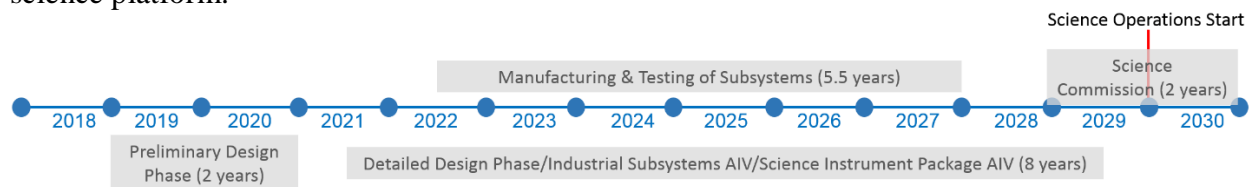

Figure 4: Current timeline and duration of project development schedule. MSE science operations begin in 2030.

The development schedule is technically paced which requires strategic planning to ensure consistent resources and cash flow. The duration of the activity blocks are bottom-up consolidations of individual development schedules provided by the subsystems as part of their Conceptual Design Phase deliverables. The project development schedule is optimized by time phasing subsystems' activities in parallel and reducing the durations of sequential activities.

# 6 Cost Estimates

## 6.1 Project Cost – Design and Construction

The Work Breakdown Structure (WBS) in Figure 5 shows the lowest level work -components used to estimate the cost of the MSE Observatory. A companion WBS dictionary facilitates the estimation process by delineating the included and excluded work in each work component with respect to its deliverables and interface responsibilities.

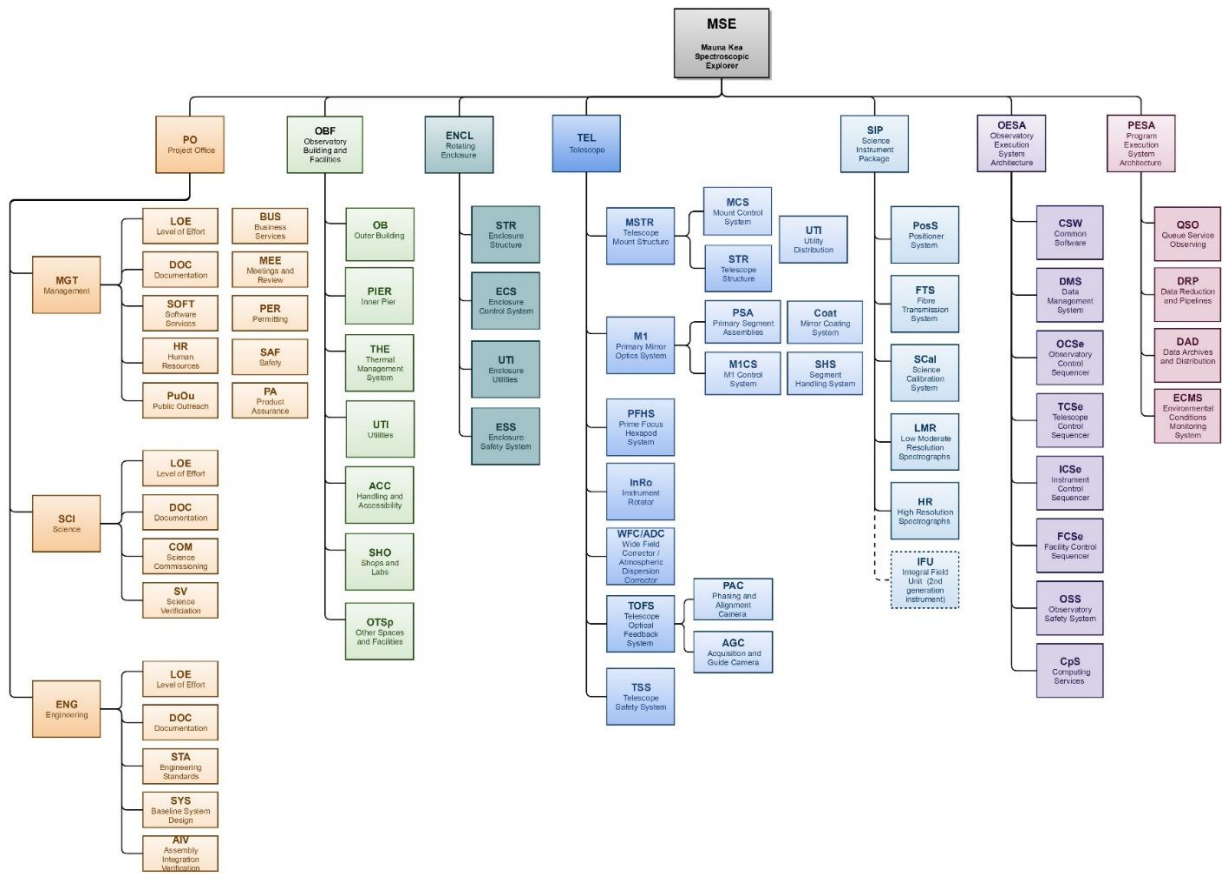

Figure 5: MSE Work Breakdown Structure and its work components

The rolled-up costs of the seven top level work components organized by their development phases are tabulated in Table 2. The seven work components are the Project Office, Observatory Building, Enclosure, Telescope, Science Instrument Package, Observatory Execution System Architecture (OESA) that executes the observations, and the Program Execution System Architecture (PESA) that manages the end-to-end science data products from acquisition, reduction, analysis, archiving, and distribution.

Table 2: Project cost in FY2017 dollars by work-components and development phases, Conceptual Design Phase cost not shown.

| Work Component | Preliminary Design Phase | Detailed Design Phase | Manufacturing and Test & AIV | Total | % of Total |
|---|---|---|---|---|---|
| Project Office | $6,730,048 | $6,885,577 | $28,941,563 | $42,570,939 | 10.0% |
| Observatory Building | $508,469 | $1,494,675 | $23,554,585 | $26,807,729 | 6.3% |
| Enclosure | $780,000 | $2,599,553 | $48,769,421 | $52,452,974 | 12.4% |
| Telescope | $9,834,762 | $18,040,510 | $131,166,943 | $159,192,215 | 37.6% |
| Science Instrument Package | $11,671,994 | $8,061,494 | $92,940,466 | $112,673,954 | 26.6% |
| Observatory Control System (OESA) | $941,805 | $1,412,707 | $2,874,972 | $5,229,484 | 1.2% |
| Observing Program Execution Software System (PESA) | $3,750,000 | $8,750,000 | $12,500,000 | $25,000,000 | 5.9% |
| Total | $34,217,078 | $47,244,516 | $340,747,950 | $423,927,295 | 100.0% |

The total risk adjusted cost, base cost + risk cost, is $423.9M in base year FY2017. The total cost reflects the conceptual design architecture, and is the rolled-up cost of individual work component risk adjusted costs provided by the design teams at the lowest WBS level. The risk cost is 25% of the base cost, and it reflects the assessments of design maturity, technical complexity and schedule criticality of the system and subsystem designs. The design teams used standard methodology provided by the Project Office to estimate the base and risk costs. The level-of-effort costs associated with labor are tied to the durations of the development schedule.

### 6.2 Operating Cost

The MSE operating cost estimate assessed by the Project Office is $18.4M in base year FY2018, with a 44% risk cost to base cost ratio which reflects the stage of development of the operations model. The operating cost is estimated bottom-up based on the science platform aspiration described in the Project's Operations Concept Document. The staff's functionalities, skill levels and skill sets required to perform the observatory operations form the basis of the staffing cost estimate. The observatory functionalities are divided into seven operations groups, Figure 6, and each group is allocated with staff members in units of FTEs corresponding to their unique skill sets and skill levels anticipated. Based on the conceptual design architecture, a total of 69 FTEs is expected. The staffing cost is calculated using CFHT's full-up annual salary of $120K per FTE.

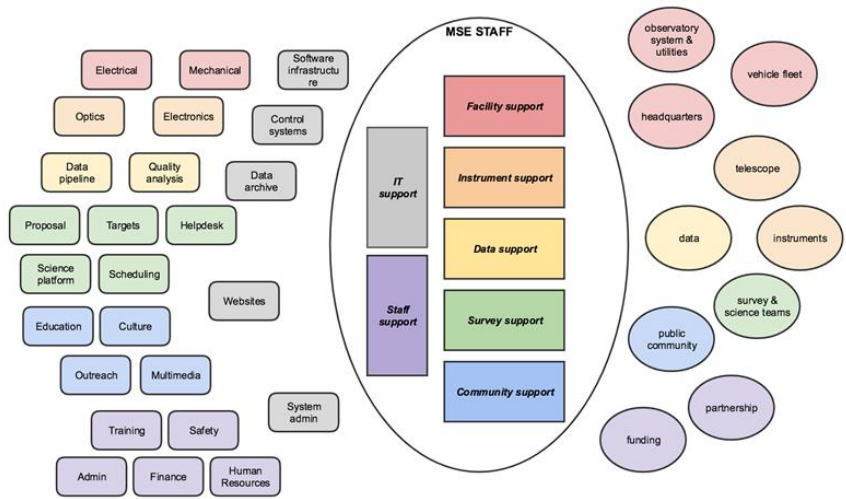

Figure 6: Color coded schematic of the MSE Staff in seven operations groups (middle), group core responsibilities (right), and functional keywords highlighting the typical functionalities of each group (left).

The non-staff cost is based on historical data from the CFHT Observatory. The non-staff cost is divided into 10 parts, ranging from consumables, utilities for the summit and headquarters, business overheads, fee for the shared Maunakea Support Services among the observatories, public outreach, sublease rent, annual allowances for future observatory upgrade, and to site restoration after decommissioning. Where appropriate, each of these is adjusted to reflect the differences in operation requirements between CFHT and MSE.

### 6.3 Partnership Funding Model

The costs for the design, construction and operations of MSE are shared by the international partners according to their fractional shares. The current partners are national astronomical institutes and universities in Australia, Canada, China, France, India and the US. The fractional share determination is based on the accumulated values of the partners' contributions, in-kind and cash, towards MSE's design, construction and operations. The Project Office tracks the contributions recognized by the governing board with representatives from the international partners. The CFHT governing board has declared MSE to be the future of the observatory and have committed to support construction and operations at the current level, as a minimum, while exploring additional resources. Moreover, the MSE partnership is evolving as more partners are anticipated. It is expected that US-based scientists will continue to comprise roughly 25% of the MSE Science Team, and therefore that the US astronomical community (i.e. universities/institutes, NSF, and/or DOE) fund roughly 25% of the project costs. Specifically, a substantial fraction of the US share of MSE construction costs is at a level appropriate for funding via NSF's mid-scale (MSIP) program: for example. In principle 1/6 of the total project cost ($70M) could be funded by MSIP (over several cycles) with supplemental in-kind contributions by US partners to reach the proposed 25% funding level by US institutions. Furthermore, the US share of MSE operations could be supported under NOAO/NCOA operations by integrating MSE's science data products into the NOAO Data Lab science platform. In MSE's current partnership model all partners have access to all data products, so this relatively small level of investment by US funding sources should result in an enormous level of science productivity for the entire US astronomical community.

## 7  Summary

MSE is a completely dedicated, massively multiplexed spectroscopic facility optimized for the design, execution, and scientific exploitation of spectroscopic surveys of the faintest objects. It will unveil the composition and dynamics of the faint Universe and impact nearly every field of astrophysics across all spatial scales, from individual stars to the largest scale structures in the Universe. MSE will provide critical follow-up for millions of faint sources found in deep imaging surveys and enables important synergies between optical/IR wide-field imaging surveys and facilities operating at other wavelengths, as well as providing essential filtering of these immensely large survey datasets to enable pointed follow-up by 30m-class telescopes,. The stand-alone science potential of MSE is transformative, but moreover the strategic importance of MSE within the international network of astronomical facilities cannot be overstated.

## 8  Acknowledgments

The Maunakea Spectroscopic Explorer Conceptual Design Phase was conducted by the MSE Project Office, which is hosted by the Canada-France-Hawaii Telescope. MSE partner organizations in Canada, France, Hawaii, Australia, China, India, and Spain all contributed to the Conceptual Design. The authors and the MSE collaboration recognize the cultural importance of the summit of Maunakea to a broad cross section of the Native Hawaiian community.